\documentclass{article}
\usepackage{appendix}

\usepackage{dcolumn}
\usepackage{bm}

\usepackage{amsthm}
\usepackage{amssymb}
\usepackage{graphicx}
\usepackage{breqn}
\usepackage{amstext}
\usepackage{amsmath}
\usepackage{amsfonts}
\usepackage{units}
\usepackage{mathrsfs}
\usepackage{hyperref}
\newtheorem{theorem}{Theorem}

\newtheorem{proposition}{Proposition}

\newtheorem{remark}{Remark}

\newtheorem{definition}{Definition}

\newcounter{mnotecount}[section]

\newcommand{\mnotex}[1]
{\protect{\stepcounter{mnotecount}}$^{\mbox{\footnotesize $\bullet$\themnotecount}}$ 
\marginpar{
\raggedright\tiny\em
$\!\!\!\!\!\!\,\bullet$\themnotecount: #1} }

\begin{document}


\title{$pp$-wave initial data}



\author{Alfonso Garc\'\i a-Parrado \\
	Departamento de Matem\'aticas, Universidad de C\'ordoba,\\
	Campus de Rabanales, 14071, C\'ordoba, Spain\\
	e-mail: agparrado@uco.es  }

\maketitle

\begin{abstract}
An {\em initial data characterization} for vacuum $pp$-wave spacetimes in dimension four
is constructed. This is a vacuum initial data set plus some extra conditions guaranteeing that
the data development is a subset of a vacuum $pp$-wave. Some of the extra conditions only depend on the 
same quantities used to construct the vacuum initial data, namely the {\em first} and the 
{\em second fundamental forms} while others are related to a {\em conformal Killing initial data characterization}
(CKID).

\end{abstract}






%
%
%
%
%
%

\section{Introduction}

An initial data characterization is a set of conditions that guarantee that an initial data set of 
the Einstein equations corresponds to a given exact solution of the Einstein's field equations. The simplest example 
are the standard {\em vacuum constraints} that guarantee that the data development is a subset of a vacuum solution.

Other more specfic examples are the construction of initial data for the Schwarzschild solution
\cite{GARVALSCH}, the construction of initial data for the Kerr solution \cite{GARKERR}, construction
of initial data for a general vacuum type D solution \cite{AGPTYPEDDATA} the construction of vacuum Killing initial 
initial data \cite{COLL77,MONCRIEF-KID,CHRUSCIEL-BEIG-KID}, its recent generalization for vacuum conformal Killing initial 
data \cite{CKID} and the initial data for the {\em closed conformal Killing-Yano equation} obtained in 
\cite{garciaparrado2020closed}. 
Some of these inital data characterizations (Schwarzschild, Kerr, type D)
share the common property of being {\em ideal} in the sense that 
they only involve the {\em natural} variables used in the set of vacuum constraints, namely, the first and second
fundamental forms. These natural variables are used in an {\em algorithmic way}, i.e., there is an algorithm to check
whether a vacuum initial data set belongs to one of the initial data characterizations mentioned before.

The purpose of this letter is to extend the previous results by providing a new vacuum initial data
characterization for the case of {\em vacuum plane fronted waves with parallel rays} or in short 
vacuum {\em $pp$-wave} solutions. 
This initial data characterization is derived from the conformal Killing initial data characterization found in
\cite{CKID}. This is because we shall take advantage of the existence of 
{\em proper} conformal Killing vectors in a vacuum $pp$-wave spacetime to find our main
result (Theorem \ref{theo:ppwave}).

As is well-known since Penrose's seminal work \cite{RevModPhys.37.215}, a $pp$-wave is 
not always globally hyperbolic. A detailed study of the global properties of a general 
$pp$-wave spacetime can be found in 
\cite{Flores2006}. In particular this reference addresses under which conditions 
a $pp$-wave is globally hyperbolic. Note that in this work we deal with {\em local}
solutions of the vacuum Einstein equations which in principle might be globally
hyperbolic, even though its maximal {\em analytic} extension is not.

The plan of this work is as follows: in Section \ref{sec:framework} we introduce our conventions,
in Section \ref{sec:ckid} we review the conformal Killing initial data characterization presented
in \cite{CKID} that is the starting point to obtain our main result. This is 
in Theorem \ref{theo:ppwave}, presented in Section \ref{sec:main-result}.

\section{Geometric framework}
\label{sec:framework}
We shall work with two manifolds: a 4-dimensional Lorentzian manifold
$(M, g_{ab})$ and a 3-dimensional Riemannian manifold $(\Sigma,h_{AB})$. 
All their structures are assumed, 
unless otherwise stated, 
to be smooth. 
In this work we 
will use abstract indices to denote tensor fields. 
The signature convention for the Lorentzian manifold is $(-,+,+,+)$. 
Small Latin indices $a,b,c,\dots$ will be used for
tensors on $M$ and capital Latin indices $A,B,C,\dots$
for tensors defined in $\Sigma$.

\begin{definition}
Let $(M, g_{ab})$ be a $4$-dimensional smooth connected Lorentzian manifold
whose Levi-Civita connection is given by $\nabla_a$
and $(\Sigma, h_{AB})$ a $3$-di\-men\-sional Riemannian manifold whose Levi-Civita connection 
is given by $D_A$. An i\-so\-metric embedding is an embedding map
$\phi:\Sigma\rightarrow \mathcal M$ such that $\phi^*g=h$.
\end{definition}
One can use the embedding map $\phi$ to define the pull-back bundles
$\phi^*(T(M))$, $\phi^*(T^*(M))$, each with base manifold $\Sigma$, and take them 
as the starting point to construct a tensor bundle
in the usual fashion. We will also use small Latin indices as the abstract indices to denote
tensors from this tensor bundle.
For example, the unit vector field $N^a$ to $\phi(\Sigma)$ can be regarded as a section of 
$\phi^*(T(M))$.
However, it is possible to extend $N^a$ smoothly off the 
hypersurface $\phi(\Sigma)$ and get a smooth section of $T(M)$ if we set up a foliation of $M$
containing $\phi(\Sigma)$ as one of its leaves and define the 
normal vector field to each leaf. 
We shall still use 
the same symbol $N^a$ for such extended vector field, leaving to the 
context to decide whether we are dealing with the extended vector field or its restriction.

The differential of the map $\phi$ induces {\em solders} $\phi_a{}^A$, $\phi_A{}^a$ 
between the vector bundles $\phi^*(T(M))$ and $T(\Sigma)$ that allow us to relate tensor
fields arising from the respective tensor bundles. 
Since $\phi$ is an embedding, the solders have the properties
\begin{equation}
 \phi_a{}^A\phi_A{}^b\stackrel{\Sigma}{=}\delta_a{}^b\;,\quad
 \phi_A{}^b\phi_b{}^B=\delta_A{}^B,
\label{eq:solder-relation}
\end{equation}
(the symbol $\stackrel{\Sigma}{=}$ means equality among tensor fields 
belonging to a tensor bundle arising from $\phi^*(T(M))$ ).
For example the relation
$\phi^*g =h$ can be written as 
\begin{equation}
	h_{AB}=\phi_A{}^a\phi_B{}^b g_{ab}\;,\quad
\end{equation}
from which we deduce using \eqref{eq:solder-relation}
\begin{equation}
 	g_{ab}\stackrel{\Sigma}{=}\phi_a{}^A\phi_b{}^B h_{AB}.
\end{equation}

\begin{definition}
If the image $\phi(\Sigma)$ of 
$\Sigma$ under the isometric embedding $\phi$ is 
a \emph{Cauchy hypersurface} of a (globally hyperbolic subset) of $M$, then 
the triple $\{(M, g_{ab}), 
(\Sigma, h_{AB}, \pi_{AB}), 
\phi\}$ 
is an 
\emph{initial data characterization} of 
$(M, g_{ab})$. In that case we say that the Riemannian manifold
$(\Sigma, h_{AB})$ is an {\em initial data set} for the 
Lorentzian manifold $(M, g_{ab})$.
\label{def:initial-data-charact}
\end{definition}

\subsection{The Cauchy problem in General Relativity}

Vacuum Einstein equations $R_{ab}=0$ can be written as a hyperbolic system 
of equations whose initial data must fulfill a set of 
\emph{\em constraints}. 
According to Definition \ref{def:initial-data-charact}
this result can be regarded as a {\em vacuum initial data characterization}
\begin{theorem}[Vacuum initial data characterisation] 
Let $(\Sigma,h_{AB})$ be a 3-di\-men\-sional Riemannian manifold and suppose that
there exists a symmetric tensor field $\pi_{AB}$ on it which satisfies
the conditions (vacuum constraints)
\begin{eqnarray}
&& r+ \pi^2-\pi^{AB}\pi_{AB}=0, \label{Hamiltonian}\\
&& D^B\pi_{AB}-D_A\pi=0, \label{Momentum}
\end{eqnarray}
where $\pi\equiv \pi^{A}_{\phantom{A}A}$. Provided that $h_{AB}$ and $\pi_{AB}$ are 
smooth there exists an iso\-me\-tric embedding $\phi$ of
$\Sigma$ into a globally hyperbolic, vacuum solution
$(M,g_{ab})$ of the Einstein field equations. The set
$(\Sigma,h_{AB}, \pi_{AB})$ is then called a vacuum initial data
set and the spacetime $(M,g_{ab})$ is the data
development. Furthermore the spacelike hypersurface
$\phi(\Sigma)$ is a Cauchy hypersurface in $M$ and $D(\phi(\Sigma))=M$.
\label{theorem:vacuum-data}
\end{theorem}
See Theorem 8.9 of \cite{CHOQUETBRUHATBOOK} for a 
formulation of this result with more general differentiability 
assumptions.

\section{CKID initial data}
\label{sec:ckid}
The existence of an isometric embedding of
a $n-1$-dimensional Riemannian manifold 
$(\Sigma,h_{AB})$ into a $n$-dimensional Lorentzian manifold 
$(M,g_{ab})$ admitting a 
\emph{conformal Killing vector} 
is addressed by the following result proven in \cite{CKID}
(we adopt the same notation and conventions as when $n=4$).
\begin{theorem}[AGP, I. Khavkine, 2019,
conformal Killing initial data (CKID)] \label{thm:ckid}
Consider a globally hyperbolic Einstein vacuum Lorentzian manifold,
$(M,g_{ab})$ of dimension $n>2$ with $R_{ab}=0$, and a Cauchy hypersurface given by the 
$n-1$ dimensional Riemannian manifold $\Sigma\subset M$ with 
Riemannian metric $h_{AB}$. 
Let $v_0$ and $v^A$ be respectively a scalar and a vector field 
on $\Sigma$ and define in terms of them the following quantities on $\Sigma$
\begin{dgroup} \label{eq:div-v}
\begin{dmath}
	u \equiv \left(D_C v^C - \pi v_0\right) \;,\label{eq:u}
\end{dmath}
\begin{dmath}
	\nabla_0 u
		\equiv \frac{1}{n-1} \pi u + \left(-D^{A}{D_{A}{v_{0}}}
			+ (\pi_{AB}\pi^{AB}) v_{0} +(D^{A}{\pi}) v_{A}\right) \;.
\end{dmath}
\end{dgroup}
Using the above, the necessary and sufficient conditions
yielding a set of \emph{conformal Killing initial data} (CKID) for $v_a$
on $\Sigma$ are given by the following differential conditions:

 \begin{dgroup} \label{eq:ckid}
\begin{dmath} \label{eq:ckid0}
	D_{A}v_{B} + D_{B}v_{A} -2 \pi_{AB} v_{0}
		- \frac{2}{n-1}g_{AB}
				u 
	= 0 \;,
\end{dmath}
\begin{dmath} \label{eq:ckid1}
	D_{B}D_{A}v_{0}
	+ (2 \pi_{AC}\pi^C{}_B - \pi^C{}_C\pi_{AB} - r_{AB}) v_{0}
	- 2\pi_{C(A} D_{B)}v^{C} - v^{C}(D_{C}\pi_{AB})
	+ \frac{1}{n-1}(u\pi_{AB}+g_{AB}\nabla_{0}u)
	= 0 \;,
\end{dmath}

\begin{dmath} \label{eq:ckidaf0AB}
	D_A D_B u 
		- \pi_{AB} (\nabla_0 u)
	= 0 ,
\end{dmath}
\begin{dmath*} \label{eq:ckidaf1AB}
	(r_{A B} + \pi\pi_{A B} -\pi_{AC}\pi^C{}_B) (\nabla_0 u) 
		-(D_{(A}{\pi_{B)C}} - D_{C}{\pi_{A B}}) D^{C} u 
	= 0 \;.
\end{dmath*}
\end{dgroup}

\label{theo:ckid}
\end{theorem}
\begin{remark}\em
As it was shown in \cite{CKID} one can find initial data characterizations for the
specializations of a conformal Killing vector: if the scalar $u$ 
defined by \eqref{eq:u} is constant in $\Sigma$ then the corresponding 
conformal Killing vector is the homothetic specialization (and if the constant is zero
then we have the Killing specialization).
\label{rem:ckid}
\end{remark}

\section{Main result}
\label{sec:main-result}
 Vacuum Einstein equations are not conformally invariant. Therefore 
 if a vacuum solution admits a \emph{proper} conformal Killing vector
 then it must belong to a very specific class as the following
 theorem proven in \cite{MONCRIEFCONFORMAL} shows.
 \begin{theorem}[Eardley, Isenberg, Mardsen and Moncrief (1986)]
  Let $(M,g)$ be a four-dimensional spacetime which satisfies the 
  vacuum Einstein equations and admits a conformal killing vector field 
  $v$. Then either
  \begin{enumerate}
   \item $(M,g)$ is everywhere locally flat.
   \item $(M,g)$ is a \emph{plane-fronted wave}.
   \item $v$ is a homothetic Killing vector field.
  \end{enumerate}
\label{theo:moncrief-ppwave}
 \end{theorem}
A plane-fronted wave is a vacuum solution $(M,g_H)$ characterized by 
the existence of a null vector field $k^a$ that is covariantly constant
$\nabla_a k^b=0$. Under these condtions, a local coordinate system 
$(u,r,x^1,x^2)$ can be introduced such that the metric $g_H$ adopts the form
 \begin{equation}
g_H=-2H(u,x)du\otimes du-du\otimes dr-dr\otimes du+{\boldsymbol\delta},
\end{equation}
where $H(u,x)=H(u,x^1,x^2)$ is a smooth function and ${\boldsymbol\delta}(x)={\boldsymbol\delta}(x^1,x^2)$ is the 
Euclidean metric in dimension 2. In these coordinates the null vector field $k^a$ is given
by $k^a=\tfrac{\partial}{\partial r}$ and the vacuum condition reads 
\begin{equation}
 \square_{\boldsymbol\delta} H(u,x)=0,
\label{eq:vacuum-ppwave}
\end{equation}
where $\square_{\boldsymbol\delta}$ is the Hodge operator computed with respect to
the metric ${\boldsymbol\delta}$.
Indeed, the solutions of the conformal Killing equation on a $pp$-wave 
background are explicitly computed in \cite{Keane_2004} generalizing previous results of 
\cite{Maartens_1991} and \cite{MONCRIEFCONFORMAL}.
 
Combining Theorem \ref{theo:moncrief-ppwave} with Theorem \ref{theo:ckid}, 
we can give an \emph{initial data characterization}
of a plane-fronted wave in dimension four. To that end, 
we need to make sure that
our vacuum solution is non-flat and has 
a conformal Killing vector 
that it is not homothetic. The necessary and sufficient 
conditions for that to happen are found next. First of 
all, we need an initial data characterization for the
flat Minkowski space-time in dimension 4.

\begin{proposition}
If $n=4$, then a vacuum initial data set $(\Sigma, h_{AB},\pi_{AB})$ 
is an initial data set of the flat Minkowski
spacetime, iff the following tensors defined on $\Sigma$ vanish 
identically
\begin{eqnarray}
&& E_{AB}\equiv r_{AB}+\pi \pi_{AB}-\pi_{AC}\pi^C{}_{B},\label{eq:electric}\\
&& B_{AB}\equiv\varepsilon^{KL}{}_{(A}D_{K}\pi_{L)B},\label{eq:magnetic}\
\end{eqnarray}
where $\varepsilon_{ABC}=\varepsilon_{[ABC]}$ is the volume form of $(\Sigma,h_{AB})$.
\label{prop:minkowski-data}
\end{proposition}
\proof 
The tensors $E_{AB}$, $B_{AB}$ are related to the space-time electric and magnetic parts
$E_{ac}\equiv W_{abcd}N^b N^d$, $B_{ac}\equiv ({}^*W)_{abcd}N^b N^d$ of the Weyl tensor $W_{abcd}$ by
(see e.g. eqns. (21a)-(21b) of \cite{GARVALSCH})
\begin{equation}
 E_{ab}\stackrel{\Sigma}{=}\phi_a{}^A\phi_b{}^BE_{AB}\;,\quad
 B_{ab}\stackrel{\Sigma}{=}\phi_a{}^A\phi_b{}^BB_{AB}.
\end{equation}
If $E_{ab}$, $B_{ab}$ vanish on $\phi(\Sigma)$ and $(M,g_{ab})$ is a vacuum solution then 
it is known \cite{BONILLA-SENOVILLA} that the 
Weyl tensor $W_{abcd}$ is zero on $D(\phi(\Sigma))$ and hence we deduce that the Riemann tensor 
$R_{abcd}$ vanish on $D(\phi(\Sigma))$. Therefore $(D(\phi(\Sigma)),g_{ab})$
is a subset of the Minkowski spacetime.
\qed
 
 \begin{theorem}[\bf Vacuum plane-fronted wave initial data characterization]
  Under the conditions of Theorem \ref{theo:ckid} for $n=4$, if the quantity 
  $u$ is non-constant, \eqref{eq:ckid} has a non-trivial solution 
  and the tensors $E_{AB}$, $B_{AB}$ do not vanish 
  identically on $\Sigma$, then the data development $D(\phi(\Sigma))$ must be
  a subset of a plane-fronted wave $(M,g_H)$ for some function $H$
  fulfilling the condition \eqref{eq:vacuum-ppwave}.
  \label{theo:ppwave}
 \end{theorem}
 \proof 
 Given that by assumption the conditions of Theorem \ref{theo:ckid} are fulfilled by 
 a non-trivial $v$, $v_A$,
 the vacuum solution $(D(\phi(\Sigma)),g)$ has a conformal Killing vector that cannot be 
 a homothetic or Killing specialization, since $u$ is non-constant (see 
 Remark \ref{rem:ckid}). Therefore 
 $(D(\phi(\Sigma)),g)$ has a {\em proper} conformal Killing vector and since $(D(\phi(\Sigma)),g)$
 is a four-dimensional vacuum solution, then the result follows from Theorem
 \ref{theo:moncrief-ppwave}.
 \qed
 
The previous result is a characterization of an isometric 
embedding of a Riemannian manifold into a four-dimensional vacuum plane-fronted wave
because $(D(\phi(\Sigma)),g)\subseteq (M, g_H)$ for some $H$ solving \eqref{eq:vacuum-ppwave}. 
The equality can only hold in those cases in which $(M, g_H)$ is globally hyperbolic.

\section{Conclusions}
An initial data characterization of a vacuum $pp$-wave spacetime has been found in Theorem \ref{theo:ppwave}.
This initial data characterization involves the first and the second fundamental forms 
$h_{AB}$, $K_{AB}$, fulfilling the vacuum constraints \eqref{Hamiltonian}-\eqref{Momentum},
the quantities $v$, $v_A$ fulfilling the CKID conditions \eqref{eq:div-v}-\eqref{eq:ckid}
and a check that \eqref{eq:electric}-\eqref{eq:magnetic}
are not identically zero. 

\section*{Acknowledgements}
We thank Dr. Igor Khavkine for a careful reading and 
his many suggestions that improved an earlier version of this manuscript. 
Funded by projects PY20-01391 and UCO-1380930  from
the Regional Government of Andalusia (Spain) and ERDEF (UE).


%



\end{document}